\documentclass[nofootinbib,floats,letterpaper,floatfix,eqsecnum,prd,aps,10pt]{revtex4}
\usepackage{dcolumn,epsfig,minitoc}
\usepackage{hyperref}
\usepackage{amssymb,amsmath}
\usepackage[usenames]{color}

\def\b{\beta}
\def\g{\gamma}
\def\n{\nu}

\def\O{\Omega}

\def\o1{$O_{1}$}
\def\o2{$O_{2}$}

\def\lb{\label}
\def\beq{\begin{equation}}
\def\eeq{\end{equation}}
\def\bea{\begin{eqnarray}}
\def\eea{\end{eqnarray}}
\begin{document}

\markboth{Mariano Cadoni}
{SCALAR BLACK BRANES WITH NON-ADS ASYMPTOTICS }

%
%

\title{SCALAR BLACK BRANES WITH NON-ADS ASYMPTOTICS\footnote{Talk 
given at the Thirteenth Marcel Grossmann Meeting, Stockholm, 1-7 july 2012.} }

\author{Mariano Cadoni}

\address{Dipartimento di Fisica, Universit\`a di Cagliari and INFN, Sezione di
Cagliari, \\
Cittadella Universitaria,
09042 Monserrato, Italy}


\begin{abstract}
Black brane solutions with scalar hair and  with non-AdS asymptotics have several 
features that make them very interesting  for holographic  
applications:
(1) They allow to circumvent usual no-hair theorems; (2) 
They may have a non singular extremal limit in the form of a 
scalar soliton that interpolates between AdS and non-AdS vacua; (3)
They allow for phase transitions between the Schwarzschild-AdS (SADS) 
and scalar dressed black branes;
(4) They give an holographic realization of 
hyperscaling violation in critical systems.
In this note I  illustrate these features  using an exact integrable Einstein-scalar 
gravity model.

\end{abstract}


\maketitle

\section{Introduction}	
In the weak field approximation (i.e. far away from the sources) the gravitational field 
generated by 
localized sources is described by  Minkowski spacetime $ds^{2}= 
-dt^{2}+ dr^{2} + dx_{i}^{2}¥$. 
In the presence of a negative cosmological constant the 
spacetime becomes Anti de Sitter (AdS) $ds^{2}= 
r^{2}(-dt^{2}+ dx_{i}^{2}¥)  + r^{-2}dr^{2} $, which is particularly 
relevant for  holographic applications, in particular for the  AdS/CFT 
correspondence.

One is therefore naturally led to consider spacetimes 
that interpolate between  Minkowskian and AdS behaviour
\beq \lb{BB}
ds^{2}= 
r^{\eta}(-dt^{2}+ dx_{i}^{2}¥)  + r^{-\eta}dr^{2}, 0\le \eta \le 2.\
\eeq
This spacetime  is obviously  not maximally symmetric and 
describes a brane,  which must be 
sourced by a non trivial scalar field   behaving as $\ln r$ 
\cite{Cadoni:2011nq}\,. 
Nevertheless, spacetimes of the form (\ref{BB}) have recently attract 
much attention  because they play an important role in holographic applications, in 
particular for   the AdS/condensed matter (CM)  correspondence.

Simply putting  in the AdS background a black brane (BB)  with a non trivial scalar field 
and eventually a finite Electromagnetic  charge  gives a rich phenomenology in the dual QFT:
phase transitions triggered by  scalar condensates 
\cite{Cadoni:2009xm,Cadoni:2011kv,Cadoni:2013hna}\,,
$U(1)$  symmetry  breaking (superconductivity) 
\cite{Hartnoll:2008vx,Hartnoll:2008kx}\,,
non-trivial transport properties of the dual QFT 
\cite{Goldstein:2009cv, 
Cadoni:2009xm,Charmousis:2009xr,Cadoni:2011kv}\, . 
In a wide class of models the near-horizon behavior, corresponding 
to the infrared (IR) regime of the dual QFT,  is described by the 
metric (\ref{BB}) or by its Lifshitz generalization 
\cite{Charmousis:2009xr,Cadoni:2011nq,Dong:2012se}
\begin{equation}
ds^{2}=r^{\theta -2}(-r^{-2(z-1)}dt^{2}+dx_{i}^{2}+dr^{2}),  \label{hv}
\end{equation}
where $\theta $ is the hyperscaling violation parameter and $z$ the dynamic
critical exponent. 

Although most of the conformal isometries of the AdS spacetime are broken 
whenever $\eta\neq 2$, the backgrounds (\ref{BB}) and (\ref{hv}) still preserve some 
scaling symmetries, namely they are  {\sl scale covariant}
\cite{Dong:2012se,Cadoni:2011nq}\,. In the dual QFT  
this corresponds to hyperscaling 
violation. 

Most of the models  with hyperscaling 
violation in the IR investigated in the literature  are 
low-energy effective  models, which flow in the  ultraviolet (UV) 
to an AdS spacetime. However, we will show   that an UV completion of 
the models is not strictly necessary. In fact all the thermodynamical 
parameters and properties are well defined also for models with 
hyperscaling violation in the UV. Moreover, we will show that a wide 
class of models exist which have an AdS phase in the IR and flow to 
an hyperscaling violating phase in the UV.

This note is mainly based  on the  
papers \cite{Cadoni:2011yj,Cadoni:2012uf}\,. Its structure is the following. In sect. \ref{I} we 
will describe the BB solutions of Einstein-scalar 
gravity with non-AdS asymptotics. In Sect. 
\ref{II} we will investigate their thermodynamics and  phase 
transitions.  In Sect. \ref{III} we will be concerned with the 
symmetries of the solutions. 
\section{Black branes of  AdS Einstein-scalar gravity}
\lb{I}
We consider  a model of  minimally coupled Einstein-scalar gravity in 
4D \cite{Cadoni:2011yj}\,: 
\beq\label{lagr}
S=\int d^{4}x \sqrt{-g}\left[R-2 (\partial \phi)^{2} 
+\frac{6}{\gamma L^{2}}\left(e^{2\sqrt3\beta 
\phi}-\beta^{2}e^{\frac{2\sqrt3}{\beta}\phi}\right)\right],
\eeq
where L is the AdS  length, $\beta$ is a free-parameter and 
$\gamma=1-\beta^2$. Note that the action is invariant under duality 
transformation $\beta \to 1/\beta$.

The model has several interesting features \cite{Cadoni:2011yj}\,.
It  is a fake supergravity (SUGRA) model 
\cite{Freedman:2003ax,Townsend:1984iu}\,. For the asymptotically AdS 
solutions holds a positivity energy theorem  and related no-hair 
theorem  forbidding  hairy  BB solutions.
The field equations for static, planar, radially symmetric solutions are exactly integrable 
(reduce to those of a Toda molecule).
The model allows  for BB solutions with  
scale-covariant asymptotics and it  emerges as Kaluza-Klein compactifications 
of $p$-brane solutions of SUGRA theories.

The field equations stemming from Eq. (\ref{lagr}) can be reduced to that of a Toda Molecule and
admit   
a two-parameters ($\n_{1},\n_{2}$) family of  scalar BB solutions \cite{Cadoni:2011yj}\,, which  
asymptotically approach to a scale-covariant metric (\ref{hv}) with 
exponents: $z=1,\theta =\frac{6\beta ^{2}}{3\beta 
^{2}-1}$.
The extremal limit  is a  completely regular scalar soliton 
interpolating between  AdS in the near-horizon (IR) region and a scale covariant  metric 
asymptotically (UV) region.
In the IR limit conformal invariance is restored. This corresponds to an IR fixed point of the dual QFT.
In the UV  the metric takes the form (\ref{BB}).  This corresponds to hyperscaling violation in the dual QFT.
Apart from the scalar BB the model obviously admits the SADS solution 
with $\phi=0$. 

\section{Thermodynamics  and phase transition}
\lb{II}

Thermodynamics of the two-parameter family of  BB solutions  of Ref. 
\cite{Cadoni:2011yj}
is not straightforward. Temperature and entropy density  are given by 
standard  formulae \cite{Cadoni:2012uf}
\beq\lb{TS}
T=\frac{1}{4\pi}\frac{3\g}{1+3\b^{2}}(\n_{1}+\n_{2})^{-\frac{2\b^{2}}
{3\g}}\n_{1}^{1/3}, \qquad S= \frac{\O R^{2}}{4G^2}= 4\pi \O
(\n_{1}+\n_{2})^{\frac{2\b^{2}}{3\g}}\n_{1}^{2/3},
\eeq
where $\O$   is the volume
of the transverse 2-dimensional space.
On the other hand the  energy  density (BB mass  density) of the 
solution   diverges.
This is  due to the  explicit  dependence of the scalar from $\n_{1}$ 
(the temperature).  The 
problem can be solved imposing  the constraint 
$\n_{2}(\n_{2}+\n_{1})=1$. Using an Euclidean action formalism
we get the BB mass and free energy \cite{Cadoni:2012uf}\,:
$M= \frac{2\O}{1+3\b^{2}}\left
(\frac{1}{\n_{2}}+(2\b^{2}-1)\n_{2}\right),\, F_{SB}(T)= 
\frac{\Omega}{1+3\b^{2}}\left(\frac{3\b^{2}-1}{\n_{2}}+
(\b^{2}+1)\n_{2}\right)$. One  can easily check that  the first 
 principle  $dM=TdS$ is satisfied.
Comparing    the free energy of the scalar BB with that of the SADS  
brane one finds that 
 exists a critical temperature $T_{c}$ above which the system undergoes a 
first order phase transition between  SADS  and the scalar BB 
\cite{Cadoni:2012uf}\,.

\section{Hyperscaling violation}
\lb{III}

The  features discussed in the previous sections  have simple interpretation in 
terms of symmetries of the  solution in the UV and IR region.
In the IR phase we have AdS$_{4}$, hence full conformal symmetry  
and  critical parameters $z=1,\theta=0$.  
Conversely in the  UV phase we have the scalar BB, hyperscaling 
violation and critical parameters  $z=1,\theta=6\b^{2}/(3\b^{2}-1)$.
The  problem is that for  $\b^{2}<1/3$  the hyperscaling violation 
parameter $\theta$ is negative, whereas for  $\b^{2}>1/3$   we have 
$\theta>2$.
It has been shown that 
both regions  $\theta <0$ and $\theta >2$ are compatible with the 
null energy conditions for the bulk stress-energy tensor 
\cite{Dong:2012se,Cadoni:2012uf}\,.
However, in the region $\theta >2$  the scalar BB phase is unstable.
This instability  is confirmed by our results for the free energy and 
for specific heat. On the other hand  negative values of  $\theta$ 
are not common in CM physics. In fact they correspond to a ÒrisingÓ of the effective 
dimensions of the dual QFT \cite{Cadoni:2012uf}\,.
 This fact  is probably related to an other main difference between 
 the two cases. 
In CM physics the hyperscaling-violating phase is  stable at small $T$ where 
long wavelength fluctuations (for instance induced by a random field ) dominate 
over thermal fluctuations.
In our case the opposite is true: the hyperscaling violating phase 
is stable   at large $T$.
This interchange of UV and IR physics is a rather puzzling point.

In this note we have considered the four-dimensional case, but our 
results can be also extended to  arbitrary 
dimensions\cite{Cadoni:2012ea}\,.


\end{document}